\documentclass[twocolumn]{aastex701}
\usepackage[T1]{fontenc}
\usepackage{amsmath,amssymb,bm}
\usepackage{tikz}
\newcommand{\vdag}{(v)^\dagger}
\newcommand\aastex{AAS\TeX}
\newcommand\latex{La\TeX}
\newcommand{\psp}{\emph{PSP}}
\renewcommand{\sc}{\rm sc}
\renewcommand{\sup}[1]{^{(#1)}}
\newcommand{\pder}[2]{\frac{\partial{#1}}{\partial{#2}}}
\newcommand{\ket}[1]{\left|#1\right\rangle}
\newcommand{\bra}[1]{\left\langle#1\right|}
\renewcommand{\vec}[1]{\bm{#1}}
\newcommand{\rot}[1]{\vec\nabla\times#1}
\renewcommand{\div}[1]{\vec\nabla\cdot#1}
\newcommand{\mo}{\mu_0}
\newcommand{\er}{\vec{\hat e_r}}
\newcommand{\ephi}{\vec e_\phi}
\newcommand{\ex}{\vec{\hat e}_x}
\newcommand{\ey}{\vec{\hat e}_y}
\newcommand{\ez}{\vec{\hat e}_z}
\newcommand{\eb}{\vec{\hat b}}
\newcommand{\eref}[1]{(\ref{#1})}
\newcommand{\vphi}{\varphi}
\newcommand{\mr}[1]{\mathrm{#1}}
\newcommand{\sinc}[1]{{\rm sinc}#1}
\newcommand{\etal}{\emph{et al.}}
\renewcommand{\(}{\left(}
\renewcommand{\)}{\right)}
\renewcommand{\[}{\left[}
\renewcommand{\]}{\right]}
\newcommand{\aave}[1]{\left\langle #1\right\rangle} 
\newcommand{\eave}[1]{{\rm E}\left\{ #1\right\}}
\newcommand{\enave}[1]{{\rm E}_{\mathcal N}\left\{ #1\right\}}
\newcommand{\Ci}[1]{{\rm Ci}#1}
\newcommand{\Si}[1]{{\rm Si}#1}
\newcommand{\wind}{\emph{WIND}}

\begin{document}

\title{The integral and correlation scales of solar wind turbulence}

\author[orcid=0000-0002-8841-6443]{Jean C. Perez}
\affiliation{Department of Aerospace, Physics and Space Sciences, Florida Institute of Technology, \\
 150 W. University blvd, Melbourne, Florida, 32901, United States of America.}
\email[show]{jcperez@fit.edu}  

\author[orcid=0000-0002-2358-6628]{Sofiane Bourouaine}
\affiliation{Department of Aerospace, Physics and Space Sciences, Florida Institute of Technology, \\
 150 W. University blvd, Melbourne, Florida, 32901, United States of America.}
\email{sbourouaine@fit.edu}

\author[orcid=0000-0001-5012-0877]{Mason Dorseth} 
\affiliation{NRL address is Plasma Physics Division, Naval Research Laboratory, Washington DC 20375}
\email{dorseth2018@my.fit.edu}






\begin{abstract}

Many works have attempted to estimate the correlation and integral timescales associated with turbulent fluctuations in the solar wind, which are interpreted as length scales based on Taylor's~Hypothesis. However, accurate estimates of these timescales from spacecraft observations heavily rely on the accurate estimation of autocorrelation functions (ACF), which have been recently shown to depend strongly on the interval length used to estimate them. In this Letter, we show that this dependence on interval length may be artificial because common ACF estimators do not correctly capture the long-lag behavior of the true ACF of the underlying turbulence. We introduce a new ergodicity-based methodology to unambiguously estimate the integral timescale, and a new ACF estimator with better ergodic convergence than current ones. Due to its ergodic properties, the new ACF estimator properly captures the long-lag behavior, and is independent of the interval length. We use this approach to estimate the integral and correlation scales of magnetic fluctuations in the solar wind near $1~{\rm au}$.

\end{abstract}

\keywords{\uat{Solar Wind}{1534} --- \uat{Interplanetary Turbulence}{830} --- \uat{Solar Physics}{1476} --- \uat{Alfv\'en waves}{23} --- \uat{Magnetohydrodynamics}{1964} --- \uat{Two-point correlation}{1951}}

\section{Introduction}  \label{sec:intro}
For decades, spacecraft measurements have consistently shown that the solar wind exhibits turbulent fluctuations in plasma and magnetic field properties over a broad range of scales, making it a premier natural laboratory for plasma turbulence~\citep{bruno13}. These fluctuations are predominantly Alfv\'enic, as revealed by the high degree of correlation between the velocity and magnetic fields, and low plasma and magnetic compressibility~\citep{damicis15,damicis19,dorseth24a}. These Alfv\'enic fluctuations were shown to carry enough energy to heat the solar wind plasma \citep{hellinger11,bandyopadhyay2023,bourouaine24}. 

One of the most important characteristic scales in turbulence theory is the integral length scale and the correlation length, which are defined in the context of stationary and homogeneous turbulence~\citep{frisch95,biskamp03}. Solar wind observations have long been used to infer these scales and many other multi-scale turbulence properties from temporal measurements of plasma and magnetic field fluctuations~\citep{verscharen19}. Due to the high solar wind speed, these temporal signals arising from spacecraft measurements are interpreted as spatial variations by invoking Taylor's hypothesis (TH) of frozen-in flow~\citep{taylor38}. These turbulence characteristic lengths have been estimated as timescales associated with temporal signals from single-spacecraft measurements via TH~\citep{matthaeus10,ruiz14}, as well as through measurements from multiple spacecraft~\citep{matthaeus05,weygand13}. 

A fundamental assumption in most analyses of spacecraft observations is that turbulent signals are stationary, which is almost impossible to avoid when comparing with turbulence theories~\citep{chen16}. The validity of this assumption has been investigated by several works since~\cite{matthaeus82b}. More recently, \cite{isaacs15} and \cite{jagarlamudi19} showed that estimates of the autocorrelation function (ACF) of magnetic fluctuations in the solar wind exhibit a strong dependence on the choice of interval length, and do not seem to converge to a fixed function for sufficiently long interval lengths. One possible explanation for these results, as the authors suggest, is that solar wind turbulence may be inherently non-stationary. 

An important aspect that is often not addressed in the estimation of ACF of solar wind observations is the role of ergodicity~\citep{frisch95}. Ergodicity is explicitly or implicitly invoked when temporal averages from finite-length intervals are used to estimate true averages associated with the intrinsic statistical properties of the underlying turbulence. For stationary turbulence, ergodicity implies that estimated averages converge to their real true values when the interval length $T$ is much larger than the integral scale $\tau_{\rm int}$. The main challenge in establishing ergodicity, not previously explicitly addressed in solar wind observations, is that an accurate estimate of the integral timescale relies on ACF estimates, which, in turn, are obtained under the assumption of ergodicity.

In this Letter we address, for the first time, the role of ergodicity in the estimation of the ACF, and the integral and correlation scales in solar wind turbulence. We maintain the working assumption that solar wind turbulence is approximately stationary in the wide sense, i.e., at least the first- and second-order statistical moments (the mean, variance, and ACF) are time-independent~\citep{alessio16}. Our ergodicity tests will be based on a new methodology to estimate the integral timescale that does not require an ACF. We also introduce a new non-standard ACF estimator that is not only more accurate than commonly used estimators, such as~\cite{blackman58}, but also independent of the interval length. We use this estimator to obtain the ACFs of turbulent magnetic fluctuations in the solar wind and, in turn, unambiguously estimate important turbulence properties, such as correlation length scales of solar wind turbulence. 

This work is organized as follows. Section~\ref{sec:review} revisits the essential elements of standard correlation and spectral analysis to provide the context and notation for this work. Section~\ref{sec:method} introduces our new methodology to estimate the integral scale and the ACF, which we apply in section~\ref{sec:data} to analyze data from the~\wind~mission. In section~\ref{sec:conclusion} we present our discussion and conclusions.

\section{Correlation and spectral analysis of stationary turbulence\label{sec:review}}

\subsection{Autocorrelation and Power Spectral Density}
For simplicity, we consider a stationary random process described by a scalar random function $x(t)$, which may represent any solar wind parameter, such as any component of the magnetic field or the plasma bulk velocity. The ACF associated with $x(t)$ is defined as
\begin{equation}
    C(\tau)\equiv\eave{\tilde x(t)\tilde x(t+\tau)},\label{eq:ACFdef}
\end{equation}
where $\tilde x(t)=x(t)-\overline x$ is the fluctuating part (zero-mean component), $\overline x\equiv\eave{x(t)}$ is the mean, and $\eave{\cdots}$ represents an ensemble average over an infinite number of realizations of the random process $x(t)$. We make the weaker assumption that $x(t)$ is wide-sense stationary (WSS)~\citep{papoulis02}, i.e., $\overline{x}$ and the ACF defined in equation~\eqref{eq:ACFdef} are solely dependent on the time $\tau$ elapsed between any two measurements. The ACF $C(\tau)$ is expected to vanish in the limit $\tau\rightarrow\pm\infty$, since two measurements must be fully decorrelated when they are separated by an infinitely long time. 

The normalized ACF is then defined as
\begin{equation}
    R(\tau) = \frac{C(\tau)}{\sigma^2},\label{eq:Rdef}
\end{equation}
where $\sigma^2=C(0)=\eave{[x(t)-\overline{x}]^2}$ is the variance of turbulent fluctuations described by $x(t)$.
The Power Spectral Density (PSD) associated with $C(\tau)$ is defined in theory via the Fourier transform
\begin{equation}
    P(\omega)\equiv\frac 1{2\pi}\int_{-\infty}^\infty C(\tau)e^{-i\omega\tau}d\tau.\label{eq:PSDdef}
\end{equation}

\subsection{ACF and PSD for finite-length signals}
While in theory we assume that solar wind properties are described by a WSS process $x(t)$ defined at all times $t$,  in practice, any given realization arising from spacecraft observations is intrinsically of a finite length $T$. In this case, we can only define the ACF associated with the truncated process $x_T(t) = W_T(t)x(t)$, where $W_T(t)$ is a squared window function that is equal to 1 for $t\in[0,T]$ and zero elsewhere. The finite-length PSD of this signal over the time period $T$ is defined  as
\begin{equation}
    P_T(\omega) = \frac{2\pi}TE\left\{\left|X_T(\omega)\right|\right\}^2,\label{eq:PSDT-def}
\end{equation}
where
\begin{equation}
    X_T(\omega)=\frac 1{2\pi}\int_0^T\tilde x_T(t)e^{-i\omega t}dt\label{eq:XFT}
\end{equation}
is the finite-length Fourier transform of the centered interval $\tilde x_T(t)=x_T(t)-\overline{x}$.   
It can then be shown that
\begin{align}
    P_T(\omega) &=\frac 1{2\pi} \int_{-T}^T C_T(\tau)e^{-i\omega\tau}d\tau,\label{eq:PT_vs_CT}
\end{align}
where
\begin{equation}
C_T(\tau) = C(\tau)\left(1-\frac{|\tau|} T\right)\label{eq:CT_vs_C}
\end{equation}
is the ACF of the underlying process tapered by a triangular window in the interval $[-T,T]$. For $T\to\infty$, $C_T(\tau)$ and $P_T(\tau)$ become $C(\tau)$ and $P(\tau)$, respectively.

The ACF and PSD have been defined thus far in terms of theoretical ensemble averages; as such, these relationships hold exactly and are not subject to statistical uncertainties present in finite-sample estimators. The routine analysis of ACF and PSD from solar wind observations aims to estimate $C_T(\tau)$ and $P_T(\omega)$ based on carefully selected intervals $x_T(t)$ of length $T$. However, the interpretation of the estimated ACF requires two important considerations: i) the estimated $C_T(\tau)$ is subject to intrinsic statistical uncertainties, which can be easily reduced using well-known estimation methods with a sufficiently large sample~\citep{alessio16}; and ii) it has been shown recently that the estimates of $C_T(\tau)$ in the solar wind exhibit a dependency on the interval length $T$ that does not seem to converge to a fixed function, the true ACF, as $T$ is increased~\citep{isaacs15,jagarlamudi19}. This result is contrary to the expectation from equation~\eqref{eq:CT_vs_C} that $C_T(\tau)$ should approach the true ACF, $C(\tau)$, in the limit $T\rightarrow\infty$. 

The lack of convergence of $C_T(\tau)$ to the true ACF poses serious implications in the calculation of important turbulence characteristics derived from it, such as the turbulence integral and correlation timescales, which are the main focus of this work. As we show later, one of the main reasons most estimators of $C_T(\tau)$ do not converge to the true ACF $C(\tau)$ as $T$ increases is due to the use of poor ergodic estimates of the mean value $\overline x$ to center the signal $x(t)$, which has a direct impact in the long-lag behavior of the estimated ACF.

\subsection{Ergodicity}
In practice, estimates of the ACF are obtained by replacing ensemble averages with averages over a finite statistical sample, which are inherently subject to statistical uncertainties. The ergodic theorem establishes that the simple temporal average, or mean, of $x(t)$~\citep{frisch95}
\begin{equation}
\overline{x}_T = \frac 1T\int_0^Tx(t)dt
\end{equation}
converges to the true average $\overline{x}\equiv E\{x(t)\}$ in the limit when $T\rightarrow\infty$, i.e., $\overline{x} = \lim_{T\rightarrow\infty}\overline{x}_T$, or equivalently
\begin{equation}
E\{x(t)\} = \lim_{T\rightarrow\infty} \frac 1T\int_0^Tx(t)dt.
\end{equation}

Intuitively, this theorem suggests that if the process $x(t)$ is observed for a sufficiently long time, it will have sampled nearly all its possible values, which raises the questions of when $T$ is large enough, and what is the deviation of $\overline{x}_T$ from $\overline{x}$ as a function of $T$? An answer to these questions is of fundamental importance because, as we will show later, a poor estimate of the mean field may strongly affect the accuracy of ACF estimation. 

The average $\overline{x}_T$ is itself a random variable because it is defined as the temporal average of a single realization $x(t)$, and therefore, it deviates from the true mean $\overline{x}$ from realization to realization. However,  $\overline{x}_T$ is an unbiased estimate in the sense that its ensemble average is $\overline x$, i.e., $E\{\overline{x}_T\} = \overline{x}$. The degree of deviation of $\overline x_T$ from $\overline x$ can be quantified by the variance of $\overline{x}_T$~\citep{frisch95}
\begin{equation}
    {\rm Var}(\overline{x}_T)=\eave{(\overline x_T-\overline x)^2}\le 2\sigma^2\frac{\tau_{\rm int}}T,\label{eq:Var_vs_Tint}
\end{equation}
where $\tau_{\rm int}$ is the integral timescale defined as
\begin{equation}
    \tau_{\rm int}\equiv \int_0^\infty R(\tau)d\tau.\label{eq:Tint}
\end{equation}
Hereafter, we call ${\rm Var}(\overline x_T)$ the \emph{temporal-mean-variance}. Note that $\overline{x}_T$ will not exhibit large deviations from the true mean $\overline{x}$ when $T\gg \tau_{\rm int}$, which means that ergodicity can only be achieved when time averaging is made over interval lengths that are much longer than $\tau_{\rm int}$. Similar but more complex expressions exist to help establish ergodicity of the ACF~\citep{papoulis02}. 

A practical implication of lack of ergodicity in the analysis of solar wind signals is that, in general, it is often not possible to find intervals that satisfy specifically desired properties with relatively high $T$ so that $T\gg \tau_{\rm int}$, as solar wind is composed of mixed streams.

\subsection{The integral timescale versus the temporal-mean-variance}

To obtain ergodic estimates of ACF, PSD, and other quantities, it is imperative to have a good estimate of the mean $\overline x_T$ and other temporal averages, which at least require $T\gg\tau_{\rm int}$. One difficulty in estimating $\tau_{\rm int}$ is that it is normally computed via the integral in equation~\eqref{eq:Tint} using an ergodic estimate of the ACF at all lags up to the interval length $T$, as well as the mean $\overline x_T$ needed to center the signal. However, ergodicity tests require knowledge of $\tau_{\rm int}$, a priori, to test the condition $T\gg \tau_{\rm int}$. In other words, to test the ergodicity of the mean and the ACF used to determine $\tau_{\rm int}$, we need an accurate value of $\tau_{\rm int}$ to verify the condition $T\gg\tau_{\rm int}$ is satisfied. Aside from ergodic limitations, standard (biased) ACF estimators introduce a tapering window that alters their large-lag behavior and thus the accuracy of the $\tau_{\rm int}$ estimate. To avoid these problems, we introduce a new methodology, entirely based on ergodicity, to estimate $\tau_{\rm int}$ that does not need an estimate of the ACF.

This new methodology, which we discuss in detail in section~\ref{sec:method}, is based on the relationship between the temporal-mean-variance and the value of the PSD at $\omega=0$~\citep{papoulis02}. Setting $\omega=0$ in equation~\eqref{eq:PT_vs_CT} in combination with~\eqref{eq:CT_vs_C}
\begin{equation}
    \int_0^T R(\tau)\left(1-\frac\tau T\right)d\tau=\frac{\pi}{\sigma^2}P_T(0),\label{eq:FiniteT_Tint}
\end{equation}
and taking the limit $T\rightarrow\infty$, we obtain
\begin{equation}
    \tau_{\rm int} = \frac\pi{\sigma^2}P(0).\label{eq:Tint_vs_P0}
\end{equation}
This expression shows that $\tau_{\rm int}$ solely depends on the variance and the value of the PSD at $\omega=0$. Moreover, $P_T(0)$ and its limiting value for $T\to\infty$, can be related to ${\rm Var}(\overline x_T)$. By setting $\omega=0$ in equation~\eqref{eq:XFT} we obtain $X_T(0)=(T/2\pi)\left(\overline x_T-\overline x\right)$, 
showing that the zero-frequency Fourier component of $X_T(\omega)$ measures the difference between the temporal mean $\overline x_T$ and the true mean $\overline x$. Taking the first two moments of $X_T(0)$
\begin{align}
    E\{X_T(0)\} &= 0,\label{eq:XT0_ave}\\
    E\{|X_T(0)|^2\} & = \left(\frac T{2\pi}\right)^2{\rm Var}(\overline x_T),\label{eq:P0_vs_VarXT}
\end{align}
from where it follows 
\begin{equation}
 P_T(0)=\frac T{2\pi}{\rm Var}(\overline x_T)   \label{eq:PT0_vs_VarXT}
\end{equation}
and therefore
\begin{equation}
    \tau_{\rm int}=\lim_{T\to\infty}\frac T{2\sigma^2}{\rm Var}(\overline x_T).\label{eq:Tint_vs_VarXT}
\end{equation}
In section~\ref{sec:method}, we use this equation to estimate the integral scale without needing an estimate of the ACF.

\subsection{Standard ACF and PSD estimation}
Consider a finite-length digital signal $x_T(n)=x(t_n)$, where $n=0,1,\cdots N-1$, $N$ is the number of measurements, $t_n=n\delta t$, and $\delta t$ is the sampling time. This interval represents a realization of the process during a time window $T=N\delta t$, within which we estimate various statistical quantities for the interval length $T$. The most natural estimator for $C(\tau)$ from a single realization~is
\begin{align}
    \hat C_{u,r}(\ell) &= \frac 1{N_\ell}\sum_{n=0}^{N_\ell-1}\tilde x_T(n)\tilde x_T(n+\ell),\label{eq:CuN}
\end{align}
for time lags $\tau_\ell=\ell\delta t$, where $\ell=0,1,\cdots, N-1$ and $N_\ell=N-\ell$. The symbol (\textasciicircum) is used to differentiate estimated quantities from their true, ensemble-averaged counterparts. This estimator is unbiased because its ensemble average is $C(\tau)$. The most commonly used ACF estimator by far follows by replacing $C(\tau)$ with $\hat C_{u,r}(\ell)$ in equation~\eqref{eq:CT_vs_C} to obtain the following estimate of $C_T(\tau)$
\begin{align}
    \hat C_{b,r}(\ell) &= \left(1-\frac{|\ell|}{N}\right)\hat C_{u,r}(\ell).\label{eq:CbN}
\end{align}

A well-known and inherent limitation of the unbiased estimator is that its accuracy decreases with increasing lag, especially when $\ell$ is close to $N$, as the number of averaged samples $N_\ell$ decreases with $\ell$ within each realization~\citep{alessio16}. The biased estimator $\hat C_{b,r}(\ell)$ is often preferred because the uncertainty involved in the estimation of large lags $\ell$ is reduced (artificially) by the tapering window $w_N(\ell)=1-|\ell|/N$. However, the downside is that it is forced to drop to zero by design, which may affect mid- and long-lag behavior.

Two additional aspects of these estimators are worth highlighting: (i) they require the interval $x(t)$ to be centered around the mean $\overline x$, which can never be known exactly, because in practice we can only provide estimates from a finite sample; and (ii) they are obtained by replacing the ensemble average in equation~\eqref{eq:ACFdef} with a temporal average within the given realization, which requires ergodicity. To address (i), and in the absence of a better estimate of the mean, the most common approach for the single-realization ACF estimators is to center the interval around its temporal (local) mean, in which case $\overline x_T$ is used in place of $\overline x$. As for (ii), because the statistics are limited to the number of samples within a given realization, the convergence of the estimators to their true value may be improved by estimating the ACF over longer intervals (based on ergodicity) or by performing an average over a large number of realizations. Increasing the interval length indefinitely is often not possible without mixing streams with different properties.

For this reason, we consider two types of averages: the temporal average within one single realization, which we call the local average, and a finite ensemble average $\enave{\cdots}$ over $\mathcal N$ subintervals, which we call the global average. Better ACF estimators are obtained when equations~\eqref{eq:CuN} and \eqref{eq:CbN} are averaged across an ensemble consisting of several subintervals, or realizations, to obtain $\hat C(\ell)=\enave{\hat C_{r}(\ell)}$, where $\hat C_{r}(\ell)$ represents either of the two estimators based on a single realization. 

While neither estimator, $\hat C_u(\ell)$ or $\hat C_b(\ell)$, accurately captures the long-lag behavior of the true ACF, they are still useful in estimating $P_T(\omega)$~\citep{jagarlamudi19}, from a discretized version of equation~\eqref{eq:PT_vs_CT} 
\begin{align}
    \hat P_T(\omega_m)=\frac 1{N\delta\omega}\sum_{\ell=-N}^{N}\hat C_b(\ell)e^{-2\pi i\ell m/N}\label{eq:PT_vs_CN}
\end{align}
sampled at the discrete frequencies $\omega_m = \delta\omega m$, where $\delta\omega = 2\pi/T$ and $m=0,\pm1,\cdots, \pm N/2$. Hence, $\hat P_T(\omega_m)$ is the Discrete Fourier Transform (DFT) of $\hat C_{b,r}(\ell)$. 

\section{Methodology\label{sec:method}}

\subsection{Integral timescale estimation from the local-mean-variance}
We propose a new methodology to estimate $\tau_{\rm int}$ based on equation~\eqref{eq:Tint_vs_VarXT}. In this methodology, we estimate the temporal-mean-variance ${\rm Var}(\overline x_T)$, which we call local-mean-variance hereafter, and the fluctuation variance $\hat \sigma_T^2$ for several values of the interval length $T$. For each $T$, the local-mean-variance is obtained by estimating the average $\eave{(\overline x_T-\overline x)^2}$ from an ensemble of $\mathcal N$ intervals of the same length $T$, where $\overline x$ is taken as the global mean $\overline x_e$, i.e., the mean of means. Similarly, the fluctuation variance $\hat\sigma_T^2$ is obtained by estimating the ensemble average $\eave{[x(t)-\overline x]^2}$ across the same ensemble. For the variance $\hat\sigma_T^2$, we have two estimates that can be used for the mean  $\overline x$, namely, the local mean $\overline x_T$ or the global mean $\overline x_e$. As we show later, the variance estimate is more reliable (ergodic) when calculated with respect to the global mean. Once both the variance and the local-mean-variance are estimated, we calculate
\begin{equation}
    \hat T_0=T\frac{{\rm Var}(\overline x_T)}{2\hat\sigma_T^2},\label{eq:T0_vs_VarXT}
\end{equation}
for several ensembles of increasing length $T$. It is important to stress that the quantity $\hat T_0$ is not the integral timescale, except in the limit $T\to\infty$, in which the relative variance ${\rm Var}(\overline x_T)/\hat\sigma_T^2$ goes to zero, and $\hat T_0\to\tau_{\rm int}$ approaches the integral scale. For this reason, there is no a priori expectation that this estimate is independent of $T$. However, an estimate of $\hat T_0$ can be obtained for increasing values of the inverval length $T$ to determine if it saturates after a certain value of $T$. Thus, finding the saturation of equation~\eqref{eq:T0_vs_VarXT} provides a non-traditional way to estimate $\tau_{\rm int}$ without an estimate of the ACF. 

As we show later, this methodology for estimating $\tau_{\rm int}$ can explain a serious limitation of single-realization ACF estimators that has received little attention: the fact that centering $x_T(t)$ around the local mean $\overline x_T$ automatically makes ${\rm Var}(\overline x_T)=0$, in which case $P_T(0)$ identically zero from equation~\eqref{eq:PT0_vs_VarXT}. This implies that any ACF estimator that centers each interval around its own mean yields a zero estimate of the PSD at $\omega=0$, which imposes the constraint that the integral of $C_T(\tau)$ over $\tau\in[0,T]$ is trivially zero, even in the limit $T\to\infty$, thereby entirely removing the integral scale from the analysis. Furthermore, for its integral over $\tau$ to vanish, the ACF estimate has to take on negative values (anti-correlate) artificially at long lags, producing a completely spurious long-lag behavior. One way to circumvent this problem is to center the signal using the global mean across all available intervals, $\overline x_e$. In this case, equations~\eqref{eq:P0_vs_VarXT} and~\eqref{eq:PT0_vs_VarXT} show that the PSD at $\omega=0$ remains non-zero, allowing for the ACF estimate to have a more natural decay at long lags.

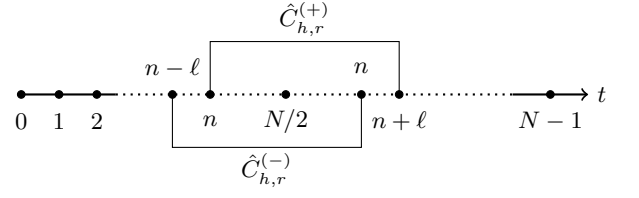
\begin{figure}
    \centering
    \begin{tikzpicture}
        \draw[thick] (-3.5,0) -- (-2.25,0);
        \draw[thick,dotted] (-2.25,0) -- (3,0);
        \draw[->,thick] (3,0) -- (4,0) node[right] {$t$};
        \foreach[count=\i] \x in {-3.5,-3,-2.5,-1.5,-1,0,1,1.5,3.5}{
            \filldraw (\x,0) circle (0.05cm);
        }
        \draw (-3.5,-0.35)     node {$0$};
        \draw (-3,-0.35)     node {$1$};
        \draw (-2.5,-0.35)     node {$2$};
        \draw (-1,-0.35)     node {$n$};
        \draw (0,-0.35)     node {$N/2$};
        \draw (1.5,-0.35)  node {$n+\ell$};
        \draw (1,0.35)  node {$n$};
        \draw (-1.5,0.35)  node {$n-\ell$};
        \draw (3.5,-0.35)  node {$N-1$};
        \draw (-1,0) -- (-1,0.7) -- (1.5,0.7) -- (1.5,0);
        \draw (1,0) -- (1,-0.7) -- (-1.5,-0.7) -- (-1.5,0);
        \draw (0.25,1)  node {$\hat C_{h,r}\sup+$};
        \draw (-0.25,-1)  node {$\hat C_{h,r}\sup-$};

    \end{tikzpicture}
    \caption{$C_{h,N}\sup +(\ell)$ (forward) and $C_{h,N}\sup -(\ell)$ (backward) ACF estimation scheme. $\tilde x_T(n)\tilde x_T(n+\ell)$ is averaged over the first half of the interval ($n=0,1,\cdots, N/2-1$) for the forward estimate, while $\tilde x_T(n)\tilde x_T(n-\ell)$ is averaged over the second half ($n=N/2,N/2+1,\cdots, N-1$) for the backward estimate.}
    \label{fig:placeholder}
\end{figure}

\subsection{New unbiased ACF estimator}
To improve the long-lag behavior of ACF estimates, we introduce a new unbiased estimator of $C(\tau)$. Each interval of length $N$ is split into two equal parts of length $N/2$ to calculate the forward and backward estimates
\begin{align}
    \hat C_{h,r}\sup+(\ell)&=\frac 2N\sum_{n=0}^{N/2-1}\tilde x_T(n)\tilde x_T(n+\ell),\\
    \hat C_{h,r}\sup-(\ell)&=\frac 2N\sum_{n=N/2}^{N-1}\tilde x_T(n)\tilde x_T(n-\ell),
\end{align}
respectively, where $\ell=0,1,\cdots,N/2-1$. The forward and backward estimates are obtained by strictly averaging over the first and second half of the interval, respectively, as illustrated in Figure~\ref{fig:placeholder}. We then define the new estimator as the simple average of these two
\begin{equation}
    \hat C_{h,r}(\ell)=\frac 12\left[\hat C_{h,r}\sup +(\ell)+\hat C_{h,r}\sup -(\ell)\right],\label{eq:ChN}
\end{equation}
which is then averaged across all available realizations
\begin{equation}
    \hat C_h(\ell)=\enave{\hat C_{h,r}(\ell)}.
\end{equation}

For this estimator, a choice of between $\overline x_T$ or $\overline x_e$ as an estimate of the mean $\overline x$ to center the intervals $\tilde x(t)=x(t)-\overline x$ is also needed. Later, we show that the estimator $\hat C_h(\ell)$ properly captures the long-lag behavior of the ACF, independent of the interval length $T$, only when the global mean is used ($\overline x=\overline x_e$).

This estimator has an important advantage over the standard unbiased ACF estimator $\hat C_u(\ell)$: the statistics for all lags contain the same number of samples, up to the largest lag $\ell=N/2$, and therefore, the statistical uncertainty is the same for all lags. Compared with the biased estimator $\hat C_b(\ell)$, the new unbiased estimator $\hat C_h(\ell)$ provides a better approximation of the true $C(\tau)$, particularly at long lags. Because it is not artificially tapered, the estimated ACF decays naturally to zero. The only trade-off with this estimator is that the maximum lag is limited to half of the available interval length.

\section{Applications: Solar wind observations} \label{sec:data}
We use high-resolution vector magnetic field data from \wind's fluxgate magnetometer~\citep{lepping95} over the 10-year period between 01/01/2013 and 12/31/2022, sampled on a uniform temporal grid with $\delta t=3~{\rm s}$ resolution. This $3628$-days-long interval provides a magnetic-field vector signal, $\vec B(m)$, of $M=104,486,400$ samples, where $m=0,1,\cdots M-1$. Assuming that magnetic fluctuations are stationary during this 10-year period, at least as a WSS process, a large statistical ensemble is obtained by splitting the 10-year-long magnetic field data into $227$ non-overlapping 16-day-long subintervals, $\vec B_T\sup s(n)=\vec B(Ns+n)$, with  $n=0,1,\cdots, N-1$, resulting in $N=460,800$ samples per subinterval. Intervals with more than 5\% of data gaps due to faulty measurements are discarded, reducing our ensemble to $\mathcal N=204$ subintervals or realizations. 

The stationarity assumption, which has been investigated extensively since the work of~\cite{matthaeus82b}, is essential and nearly unavoidable in most analyses of solar wind observations. Here, we investigate the turbulence properties of the solar wind, assuming stationarity at least approximately, and treating all fluctuations as a single turbulent system over the 10-year period. Further refinements of our analysis can be made by using a conditional analysis, which may restrict each subinterval to meet specific properties, such as fast or slow, Alfv\'enic or non-Alfv\'enic streams, to form an ensemble of intervals with similar properties. The following analysis is entirely applicable to this case, but it is not the focus of this work. Instead, we consider solar wind fluctuations as a single system, without separating them into distinct stream types.

Each of the $204$ intervals represents a 16-day-long realization from which we estimate important statistical quantities for several interval lengths $T\le 16~{\rm days}$. We use these intervals to create eight different ensembles, each one corresponding to a different interval length $T=[3, 6, 12, 24, 48, 96, 192, 384]~{\rm h}$. Each ensemble is obtained by only taking the first portion of length $T$ from each 16-day interval, for instance, if $T=6~{\rm h}$, the ensemble for this interval length corresponds to the $204$ intervals obtained from taking the first $6~{\rm h}$ from each 16-day interval. For each ensemble, we obtain estimates of $\hat T_0$, the standard biased ACF $\hat C_b(\ell)$, and the new unbiased estimate $\hat C_h(\ell)$, using the methodology described above for the following interval lengths $T=[3, 6, 12, 24, 48, 96, 192, 384]~{\rm h}$.

\begin{figure}
    \centering
    \includegraphics{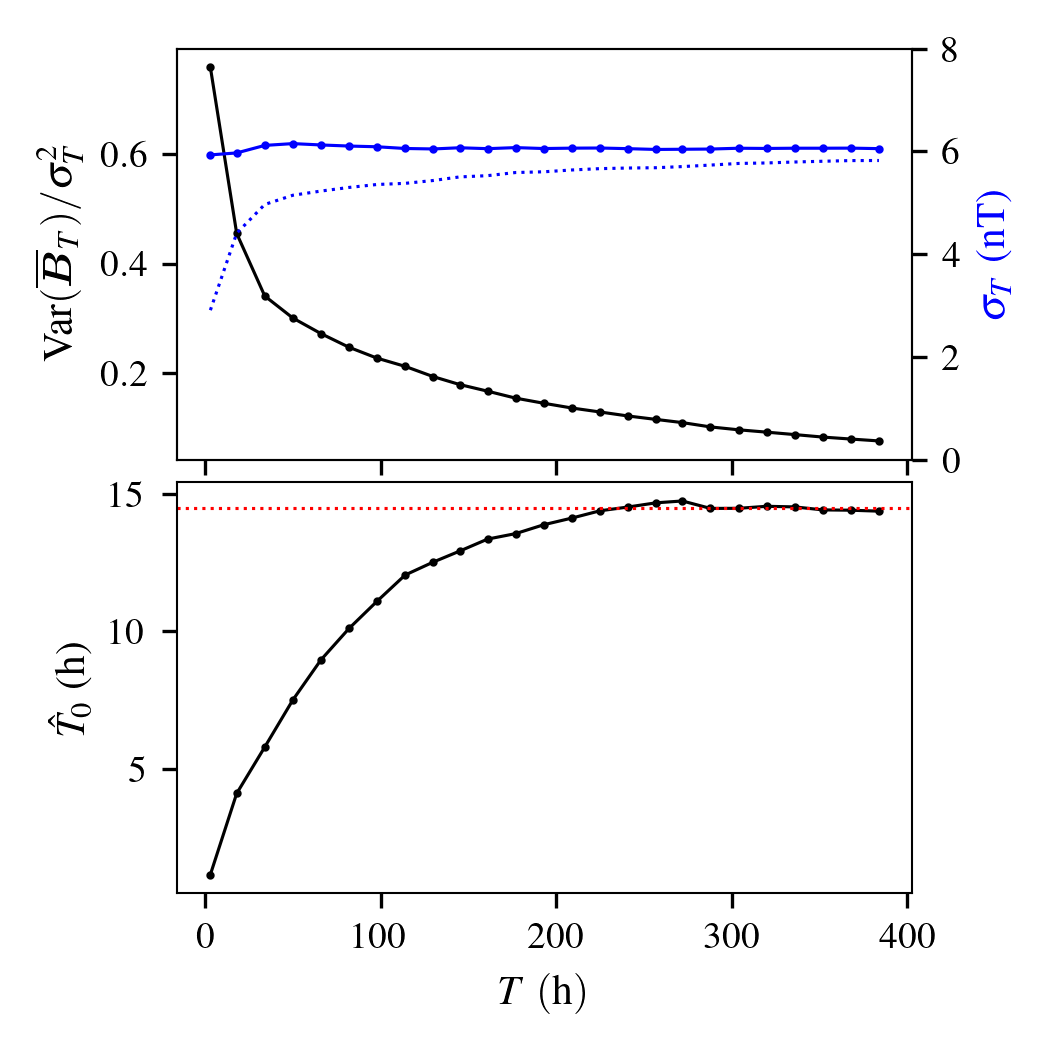}
    \caption{Top: normalized local-mean-variance (left axis), ${\rm Var}(\overline{\vec B}_T)/\hat\sigma_T^2$, and magnetic field variance (right axis), $\hat\sigma_T^2$, calculated across $204$ realizations for each time $T$ of the ensemble. Bottom: estimated $\hat T_0$ calculated for each ensemble of length $T$, saturating around $\hat T_0\simeq 14.5~{\rm h}$.}
    \label{fig:fig1}
\end{figure}

\subsection{Integral scale}
Following the methodology described above, estimates of $\hat T_0$ from equation~\eqref{eq:T0_vs_VarXT} are calculated by estimating the local-mean-variance ${\rm Var}(\overline{\vec B}_T)$ and the variance $\hat\sigma_T^2$ of magnetic fluctuations from each ensemble of interval length $T$. Each variance is calculated for each magnetic field component and combined into a total variance. We determine the integral scale $\tau_{\rm int}$ as the saturation value of the $\hat T_0$ estimate when $T$ is increased.

\begin{figure*}[!t]
    \centering
    \includegraphics{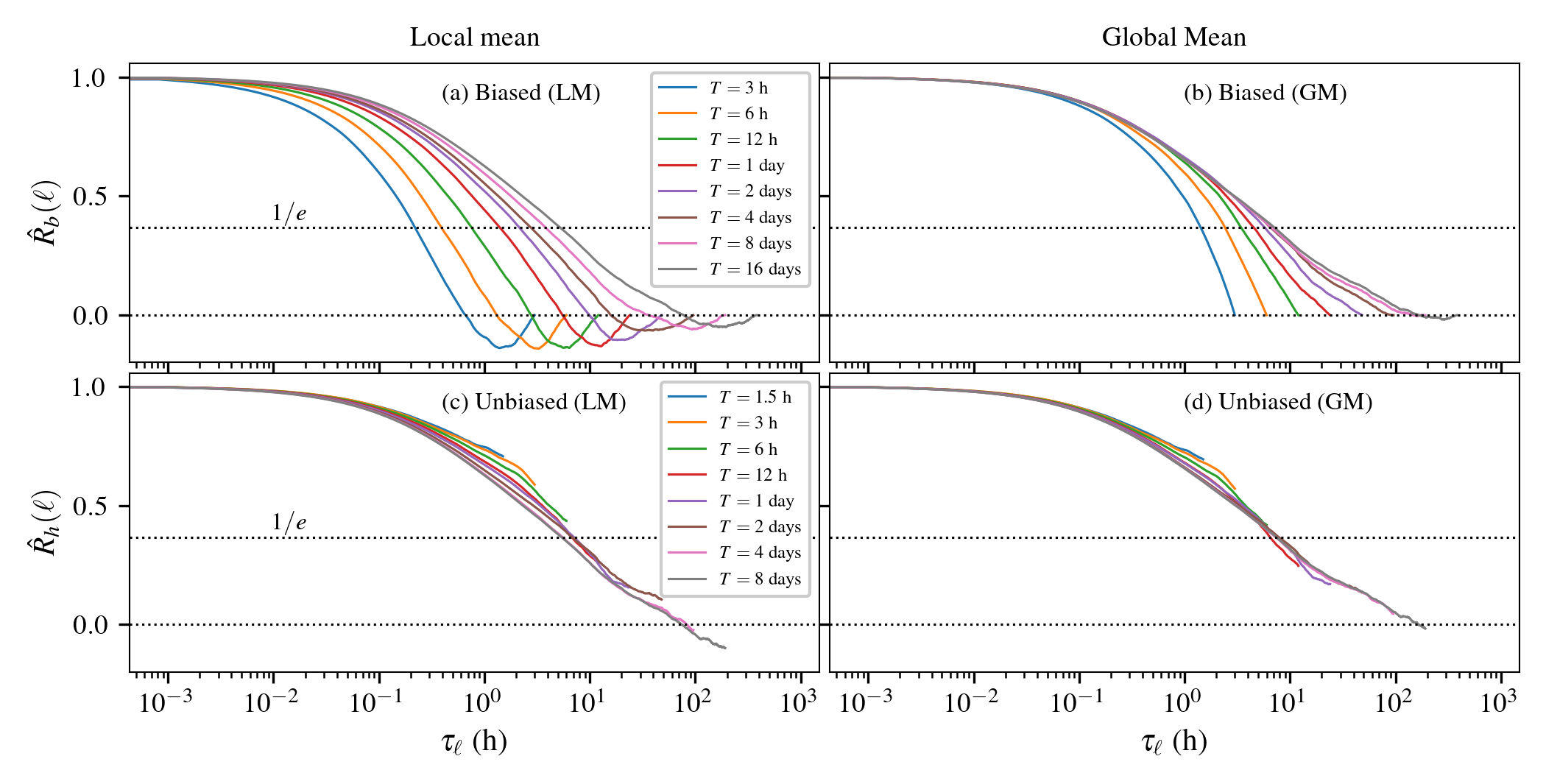}
    \caption{Estimated ACF (normalized) $\hat R_b(\ell)=\hat C_b(\ell)/\hat C_b(0)$ and $\hat R_h(\ell)=\hat C_h(\ell)/\hat C_h(0)$ vs $\tau_\ell=\ell\delta t$, with signal centered around the local mean (LM) versus global mean (GM) for different interval lengths $T$. Note that the unbiased ACF is estimated up to a maximum lag equal to half the selected interval length. Top left: Estimated $\hat R_b(\ell)$ calculated with magnetic field interval centered around its own (local) mean. Top right: Estimated $\hat R_b(\ell)$ calculated with magnetic field interval centered around the ensemble (global) mean. Bottom left: Estimated $\hat R_h(\ell)$ calculated with magnetic field interval centered around its own (local) mean. Bottom right: Estimated $\hat R_h(\ell)$ calculated with magnetic field interval centered around the ensemble (global) mean.}
    \label{fig:fig2}
\end{figure*} 
The top panel of Figure~\ref{fig:fig1} shows the estimated local-mean-variance (left black axis) normalized to the fluctuation variance (right blue axis) versus the interval length $T$ for each ensemble. As expected, the normalized local-mean-variance decreases substantially as we increase the interval length, dropping to about $7\%$ for $T=16~{\rm days}$. This result shows that to achieve this level of ergodicity in the local mean (with a single interval), time averages must be computed over 16-days. The right blue axis shows that the estimated variance exhibits very little fluctuation across $T$ when calculated with respect to the (global) mean across all intervals. At the same time, the blue dotted line shows that when the variance is estimated with respect to the local mean of each interval, it is strongly dependent on the interval length. We later observe a similar behavior for the ACF, which suggests that using the local mean may result in estimates that have not achieved ergodicity. The bottom panel of Figure~\ref{fig:fig1} shows the estimated $\hat T_0$ for each ensemble at each $T$, which exhibits saturation around $\hat\tau_{\rm int}\simeq 14.5~{\rm h}$ for $T\gtrsim10~{\rm days}$, which we identify as the integral scale.

These results show that equation~\eqref{eq:T0_vs_VarXT} allows us to estimate the integral timescale without having to estimate the ACF of the turbulent fluctuations. Moreover, this analysis shows that to obtain (local) temporal averages that do not deviate strongly from the global mean, one needs intervals of at least several days.

\subsection{ACF and PSD estimation}

Due to the vector nature of the magnetic field, we define the ACF as the trace of the correlation tensor $C_{ij}=\eave{\tilde B_i(t)\tilde B_j(t+\tau)}$. We estimate the normalized ACF, $\hat R_b(\ell)=\hat C_b(\ell)/\hat C_b(0)$ and $\hat R_h(\ell)=\hat C_h(\ell)/\hat C_h(0)$, for each of the selected interval lengths $T=[3, 6, 12, 24, 48, 96, 192, 384]~{\rm h}$. This is done by first calculating $\hat C_{b,r}(\tau_\ell),~\hat C_{h,r}(\tau_\ell)$ using equations~\eqref{eq:CbN} and~\eqref{eq:ChN}  for each of the $204$ intervals in each ensemble of length $T$, and then averaging across all the resulting ACFs. In each case, we obtain the ACF estimates by centering the signal with respect to the local (realization) mean and the global (ensemble) mean. Note that for each selected interval length $T$, the biased ACF is estimated up to a maximum lag $\tau_{\rm max}=T$ while the unbiased ACF is estimated up to $\tau_{\rm max}=T/2$.

\begin{figure*}[!t]
    \centering
    \includegraphics{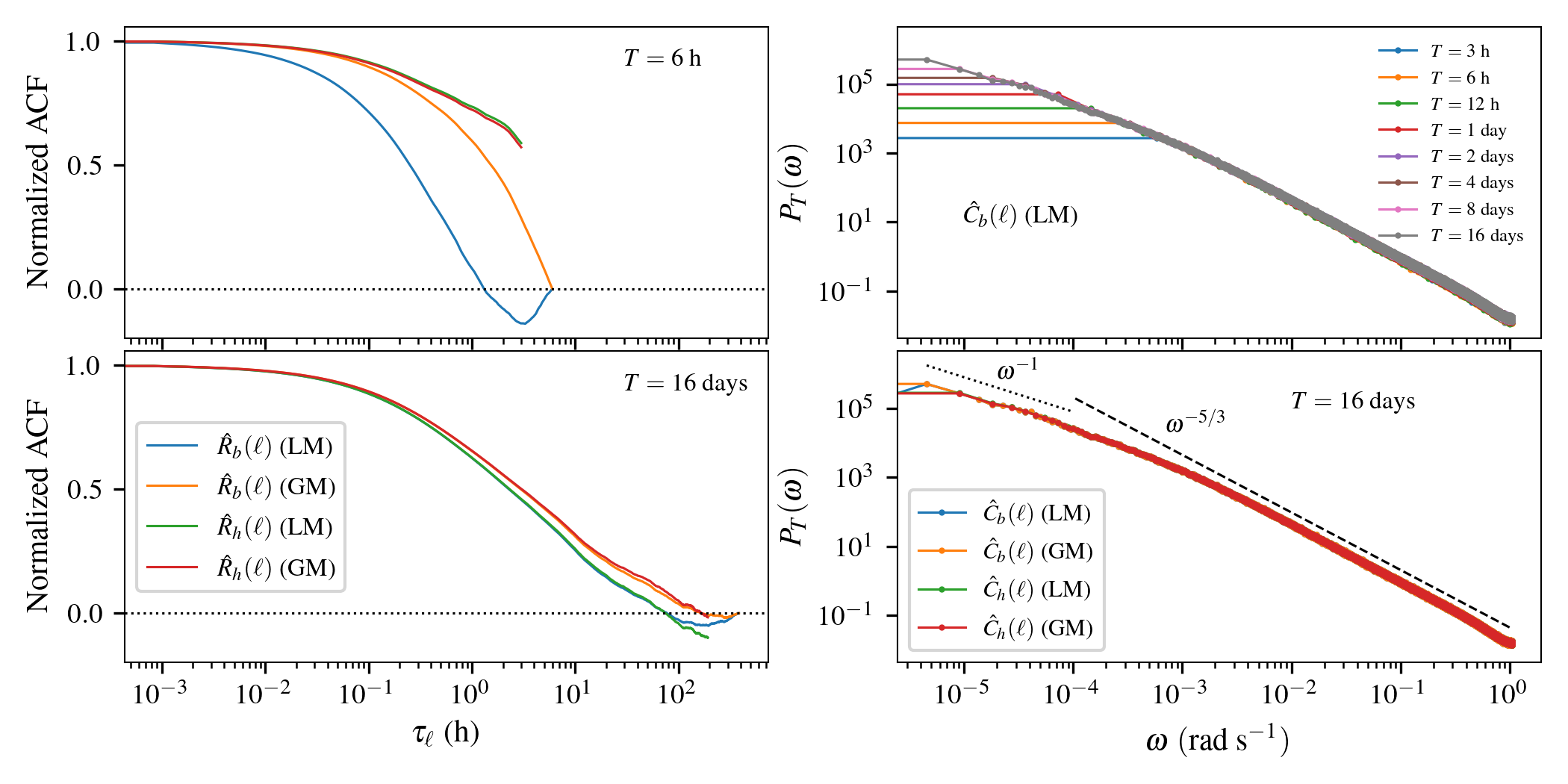}
    \caption{Left panels: comparison of biased vs unbiased ACF estimates for $T=6~{\rm h}$ (top) and $T=16~{\rm days}$ (bottom). Right panels: PSD for each ACF in Figure~\ref{fig:fig1}. The top panel shows the PSD obtained for each ACF estimate with interval lengths $T$ from 3~$ {\rm h}$ to 16 days. The bottom panel shows the PSD obtained from all ACF estimated with $T=16~{\rm days}$.}
    \label{fig:fig3}
\end{figure*}

Figure~\ref{fig:fig2} shows that the resulting ACF estimates have substantially different dependencies on the interval length $T$. The top-left panel shows that the biased estimator based on the local mean (LM) exhibits the strongest dependence on $T$, with no apparent convergence to a limiting function (for $T\rightarrow\infty$), consistent with~\cite{jagarlamudi19}. The top-right panel shows the biased estimator based on the global mean (GM), which also exhibits a strong dependence on $T$. However, this estimator appears to converge to a well-defined limiting ACF as the interval length increases beyond $T\gtrsim8~{\rm days}$.  The bottom panels of Figure~\ref{fig:fig2} show the normalized ACF estimated with our new unbiased estimator $\hat R_h(\tau_\ell)$. For these estimators, the dependence of the estimated ACF on $T$ is strongly reduced, even when the signal is centered around the local mean, and is practically nonexistent when the ACF is estimated with respect to the global mean.

Another important and known feature of the biased ACF estimate shown in the top-left panel is that it always crosses the horizontal axis ($\tau$-axis). This negative portion of the ACF is an artifact arising from the use of the local mean, which requires the ACF average to vanish, as imposed by the condition $\hat P_T(0)=0$ in equation~\eqref{eq:PT_vs_CN}. This tendency will persist for any value of $T$, as this condition forces the ACF estimate to vanish for some $\tau<T$ and become negative to produce a zero-average of $\hat R_h(\ell)$. This behavior is not present in the biased ACF based on the global mean because in this case the PSD does not vanish for $\omega=0$. However, one limitation of $\hat C_b(\ell)$ is that it is identically zero at the maximum lag $\tau=T$ by definition, from equation~\eqref{eq:CbN}. Therefore, unless $T$ is large enough that the true ACF has droppped to zero, the long-lag behavior of the estimated ACF is affected by the tapering window, which may explain the apparent dependence on $T$ in the top right panel of Figure~\ref{fig:fig2}. In contrast, in the two unbiased ACF estimates in the bottom panels of Figure~\ref{fig:fig2}, the correlation naturally drops to zero as the interval length $T$ is increased, providing a better estimate of the true normalized ACF $R(\tau)=C(\tau)/\sigma^2$.

The left panels of Figure~\ref{fig:fig3} show a comparison of biased vs unbiased ACF estimates for the shortest $T=6~{\rm h}$ (top) and the longest $T=16~{\rm days}$ (bottom) interval lengths. For the shortest intervals $T=6~{\rm h}$, the top panel shows that the biased estimate~$\hat R_b(\tau)$ based on the local mean differs the most from the other ACF estimates at all lags. The unbiased estimates remain fairly similar across all lags, regardless of the choice of magnetic field mean. In contrast, the biased estimate with the global mean remains pretty close to both unbiased estimates, $\hat R_h(\tau)$, only at very short time lags, $\tau\lesssim 10~{\rm min}$. This departure of $\hat R_b(\ell)$ with the global mean from $\hat R_h(\ell)$ at long lags is due to the tapering window present in the biased estimators. For the longest interval length, $T=16~{\rm days}$, the bottom panel shows that the biased and unbiased estimates with the global mean converge to one another, as evidenced by their overlap. Interestingly, $\hat R_b(\ell)$ and $\hat R_h(\ell)$ with the local mean also overlap for nearly all lags, except $\tau\gtrsim 60~{\rm h}$. The most notable difference among these ACFs is the long-lag behavior of those ACF estimated with the local mean vs the global mean, which, as we argue later, is mostly due to $\hat P_T(0)$.

These results indicate that the unbiased estimator $\hat R_h(\tau)$ based on the global mean is the best estimate of the ACF at any interval length because its dependence with $\tau$ has almost no variation with $T$, as shown in Figure~\ref{fig:fig2}(d), and also reproduces the (converged) biased estimate $\hat R_b(\tau)$ for $T=16~{\rm days}$ using the global mean as shown in the left bottom panel of Figure~\ref{fig:fig3}.

\begin{figure*}[!t]
    \centering
    \includegraphics{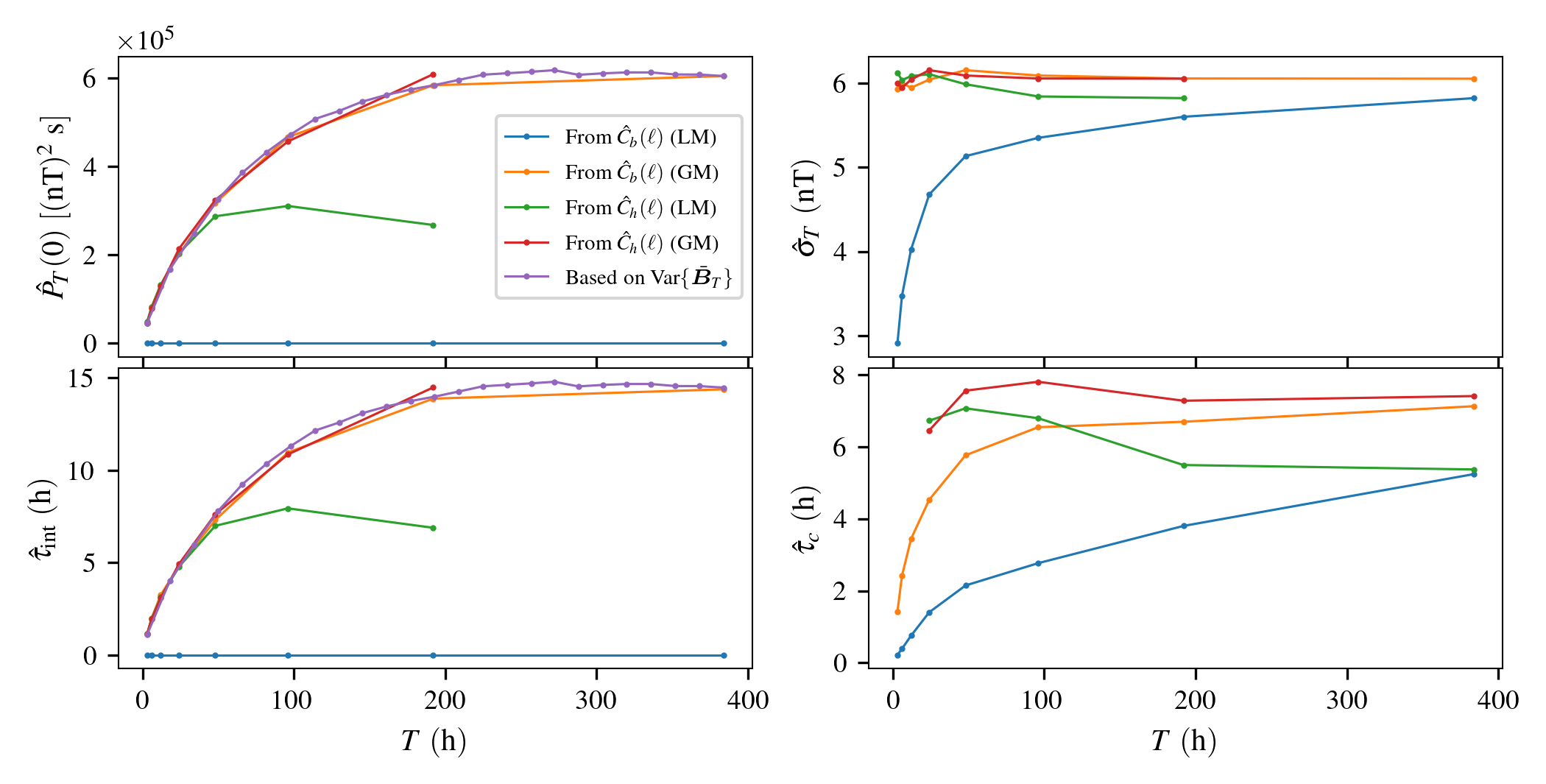}
    \caption{Top left: the estimated PSD at zero frequency for each of the four ACF estimates, and the corresponding value obtained from equation~\eqref{eq:PT0_vs_VarXT}. Top right: the fluctuation variance obtained from the ACF value at $\tau=0$ for each $T$. Bottom left: the estimated $\hat T_0$ obtained from equation~\eqref{eq:T0_vs_VarXT}. Bottom right: the correlation timescale $\hat\tau_c$ using the $1/e$ criterion for each ACF estimate. Note that for $\hat C_h(\ell)$, there is no $\tau_c$ available at lower values of $T$ where the ACF has not dropped to $1/e$.}
    \label{fig:fig4}
\end{figure*}

The right panels in Figure~\ref{fig:fig3} show the PSD obtained from various ACF estimators using equation~\eqref{eq:PT_vs_CT}. The top (right) panel shows the PSD obtained from the biased ACF $\hat C_b(\tau)$ with the local mean, which, as seen in Figure~\ref{fig:fig2}, has the strongest dependency on the interval length $T$ and shows to be the worst estimator for the long-lag dependency of the ACF. In spite of these limitations, the top right panel in Figure~\ref{fig:fig3} shows that the PSDs obtained for all $T$ display the same functional form at all frequencies above the smallest resolved frequency $\delta\omega=2\pi/T$. As the time interval length $T$ is increased, $\delta\omega$ becomes smaller, extending the PSD to smaller and smaller frequencies. This property makes this biased estimator (based on the local mean) practical for estimating the PSD at all non-zero frequencies; i.e., although the ACFs for different $T$ differ substantially, they all capture the same PSD down to the smallest frequency resolution imposed by the interval length.

These results are not surprising, and previously obtained by~\cite{jagarlamudi19}, as the effect of the tapering window on $P_T(\omega)$ from equation~\eqref{eq:PT_vs_CT} is to filter-out frequencies lower than $\delta\omega =2\pi/T$, which are associated with the long-lag behavior of $C(\tau)$. Therefore, one must be cautious when using biased ACF estimators to infer turbulence properties, such as the correlation time and the integral timescale, that depend strongly on the long-lag behavior, which is determined by the values of the PSD below $\delta\omega$. The bottom right panel of Figure~\ref{fig:fig3} shows all the PSD obtained with all estimators at $T=16~{\rm days}$, demonstrating that all ACF estimates lead to the same PSD at all non-zero frequencies. The main difference among these PSDs is not apparent in Figure~\ref{fig:fig3}, as the zero frequency value is intentionally excluded with the choice of logarithmic axes.

The behavior of $P_T(\omega)$ at frequencies smaller than the frequency resolution is entirely determined by $\hat P_T(0)$, as this is the only frequency available below $\delta\omega$. The top left panel of Figure~\ref{fig:fig4} shows $\hat P_T(0)$ for each estimated PSD as a function of the interval length $T$. As expected, $\hat P_T(0)$ for the biased ACF with local mean (LM in blue) remains practically zero (within floating-point precision) at all interval lengths $T$. For the three other PSD estimators, the value of the PSD at zero frequency exhibits similar behavior up to $T\simeq 1~{\rm day}$, after which only the two estimators using the global mean (GM in orange and red) keep increasing together until they saturate at around $P_T(0)\simeq 6\times 10^5~{\rm (nT)^2~s}$. The zero-frequency PSD for the unbiased estimator with the local mean (LM in green) saturates at a much lower value around $P_T(0)\simeq 3\times 10^5~{\rm (nT)^2~s}$. 
The top right panel of  Figure~\ref{fig:fig4} shows the root-mean-square (rms) $\hat\sigma_T=\sqrt{{\rm Var}[\vec B(t)]}$, where the variance is obtained from the value of the ACF estimate at zero lag. All but one estimators exhibit a consistent, flat value for the rms around $\hat\sigma_T\approx 6~{\rm nT}$, indicating little to no dependency on the interval length $T$. In contrast, $\hat \sigma_T$ obtained from the biased ACF with the local mean shows a strong dependency on $T$, without any sign of saturation toward an asymptotic value.

The bottom left panel shows the estimated $\hat T_0$ as a function of $T$ by computing ${\rm Var}(\overline x_T)$ directly (same as Figure~\ref{fig:fig1}), as well as from equation~\eqref{eq:PT0_vs_VarXT} using the zero-frequency value of each estimated $\hat P_T(\omega)$. Because the rms $\hat\sigma_T$ is nearly independent of the interval length $T$, except for $\hat C_b(\ell)$ with the local mean, the dependency of $\hat T_0$ with $T$ nearly mirrors the dependency of $\hat P_T(0)$. The integral timescale saturates around $\hat\tau_{\rm int}\approx 14.5~{\rm h}$ for the largest interval lengths $T\gtrsim 10~{\rm days}$. The saturation of the integral timescale based on biased and unbiased ACF using the global mean is expected for two reasons: 1) these ACF estimates are the ones who most accurately capture the long time behavior of the ACF and 2) only for $T\gtrsim 10~{\rm days}$ these ACFs drop naturally to zero (without a tapering window), as shown in Figure~\ref{fig:fig1}, to allow for the numerical evaluation of the improper integral in equation~\eqref{eq:Tint}.

Lastly, the bottom right panel of Figure~\ref{fig:fig4} shows the estimated turbulence correlation time $\hat\tau_c$, nominally defined as the time lag $\tau$ for which the normalized correlation drops to $1/e$, based on each ACF estimated for different $T$. Our results show that the unbiased ACF, $\hat C_h(\tau)$, with the global mean has almost no dependence on the interval length, providing a consistent value around $\tau_c\simeq 7.4~{\rm h}$, as opposed to the other three ACFs, which exhibit a notable dependence on $T$. For the widely used biased ACF~\citep{blackman58}, whether we use the local or global mean, the $1/e$ estimate of the correlation time shows the strongest dependence on the interval length. Using the ensemble-averaged solar wind speed across all $204$ intervals, $V_{\rm SW}\simeq 430~{\rm km~s^{-1}}$, results in a correlation lengthscale of around $\lambda_c\simeq 0.08~{\rm au}$ and an integral scale of around $\lambda_{\rm int}\simeq 0.15~{\rm au}$.

\section{Discussion and Conclusions} \label{sec:conclusion}
Most analyses of spacecraft measurements to investigate the solar wind turbulence properties assume that the turbulence is stationary. While some works have provided support for the validity of the stationarity assumption in the solar wind~\citep{matthaeus82b,perri10}, more recent works~\citep{isaacs15,jagarlamudi19} have found that the observed dependence of ACF estimated on the interval length may suggest that the solar wind is inherently non-stationary. However, this conclusion is at odds with the fact that, while the estimated ACF does depend on the interval length $T$, the estimated PSD does not, which implies that the PSD estimate is consistent with the stationarity assumption.

In this Letter, we presented the first comprehensive study on the role of ergodicity in the analysis of spacecraft measurements of solar wind turbulence. We found that the previously-reported dependence of ACF estimates on the interval length is artificial, and caused by the lack of convergence of commonly used biased ACF estimators, particularly at mid- to long-time lags. The main factors affecting the long lag behavior of standard ACF estimates are (a) the use of the local mean to center the turbulent signal and (b) biased estimators like Blackman-Tukey are designed to artificially reduce the ACF until it reaches zero at maximum lag with a tapering window, which may affect the mid and long lag behavior, particularly for estimates based on short interval lengths. Because these biased estimates do not properly capture the long lag behavior, important turbulence timescales that depend on this behavior may not be useful or lack physical meaning.

Another limitation of the standard analysis of ACF in the solar wind is that ergodic convergence of ACF estimated from temporal averaging requires interval lengths that are longer than the integral timescale $\tau_{\rm int}$, which itself is inferred from the same ACF. To overcome this, we proposed a novel methodology to estimate the integral timescale associated with turbulent fluctuations in the solar wind that does not require an ACF estimate. This methodology, which is based on the ergodicity of the temporal-mean-variance, is useful because it provides an independent means of testing whether a solar wind interval is long enough for the ergodic assumption to hold, without requiring an estimate of the ACF. Based on this methodology, we concluded that for the 10-year period we used from the~\wind~mission, the integral timescale is about $14.5~{\rm h}$, corresponding to about $0.14~{\rm au}$ using Taylor's hypothesis. We also showed that the variance of the temporal mean $\overline x_T$ substantially decreases to approximately $15\%$ of the fluctuation mean after averaging over 8-days intervals, and to about $7\%$ for 16-days intervals, consistent with the condition $T\gg\tau_{\rm int}$.

We also presented a new methodology to obtain robust estimates of the ACF of turbulent magnetic fluctuations in the solar wind. The new unbiased ACF estimator provides a more accurate estimate of the ACF behavior at long lags and independent on $T$, which is critical for obtaining important derived quantities, such as the turbulence correlation time. The robustness of this unbiased estimator of $C(\tau)$ is due to better ergodic properties: 1) the ACF estimate has the same level of statistical samples for all lags, and decays naturally to zero without tapering windows; and 2) it uses the global mean to estimate the fluctuating part of the magnetic field, thereby capturing the zero-frequency PSD $P_T(0)$, which is often ignored by local-mean-based estimators. 

Our results showed that the new unbiased estimator exhibits essentially the same functional form for all the selected time intervals from $3~{\rm h}$ to $8~{\rm days}$, from which $1/e$ correlation time of around $\tau_c\simeq 7.4~{\rm h}$ is obtained, independent of the interval length, corresponding to a correlation length of about $0.08~{\rm au}$ based on Taylor's Hypothesis. Despite the fact that the integral and correlation timescales are often treated to be the same (or at least comparable)~\citep{ruiz14}, we obtain very different numerical estimates for these timescales. The integral timescale is almost universally defined in terms of equation~\eqref{eq:Tint}, which represents a global property of the ACF, involving all scales from $\tau=0$ to $\tau\to\infty$. In contrast, the most common definition of the correlation time is a nominal time after which the ACF has decay enough so that it fully decorrelates, which is a local property of the ACF and often based on the $1/e$ criterion. Based on these definitions, there is no reason to expect these timescales to be the same, except in the unique situation where the ACF has exponential decay. In this sense, our results suggest that because these timescales are different, the ACF decay is not exponential.

The new estimator $\hat R_h(\ell)$ reproduces the same converged estimate obtained by the biased estimator $\hat R_b(\ell)$ calculated at the longest interval length $T=16~{\rm days}$ using the global mean, as shown in the bottom left panel of Figure~\ref{fig:fig3}. The standard ACF estimate, $\hat R_b(\ell)$, is observed to agree with the new ACF estimator, $\hat R_h(\ell)$, only for the largest interval length $T=16~{\rm days}$ and based on the global mean. This result is consistent with the integral timescale we obtained independently $\tau_{\rm int}$, which is much smaller than $T$, resulting in a more ergodically convergent biased ACF. The most significant advantage of our estimator over the standard one is that it can recover the converged form of the ACF without using large interval lengths $T$, as long as we construct the estimate from a large ensemble of intervals of the same length.

All the ACF estimators we tested yielded nearly identical PSD estimates, which we argue is evidence that, at least with respect to PSD, the turbulence is stationary. The PSDs differed substantially only at zero frequency, which we show is the critical factor responsible for the long-lag behavior of the estimated ACFs. We therefore conclude that our results are consistent with the stationarity hypothesis in the solar wind, at least approximately, over the 10-year window used in this analysis.

Lastly, our analysis of 10 years of \wind~data did not include conditioning to classify intervals and investigate their turbulence characteristics according to their properties, such as fast versus slow streams~\citep{wicks10,chen12}, Alfv\'enic versus non-Alfv\'enic~\citep{damicis19,dorseth24}, or sampling angle with respect to the magnetic field to account for turbulence anisotropy~\citep{horbury08,wu19,wang19}, as well as in the context of switchbacks vs non-switchbacks observed by the Parker Solar Probe mission~\citep{bourouaine20a}. It is worth noting that the use of the global mean in the estimation of conditional ACFs was first emphasized by~\cite{bourouaine20} in the analysis of Helios data. The analysis based on the methodology we presented in this Letter can be applied to investigate differences in turbulence properties across different types of solar wind streams using a conditional analysis of spacecraft observations.

\begin{acknowledgments}
JCP and SB were supported by NASA grants 80NSSC21K1768, and 80NSSC24K0137. MD was supported by NRL Base funds.
\end{acknowledgments}

\bibliographystyle{aasjournalv7}

\end{document}